\documentclass[conference]{IEEEtran}
\IEEEoverridecommandlockouts
\usepackage{cite}
\usepackage{amsmath,amssymb,amsfonts}
\usepackage{graphicx}
\usepackage{textcomp}
\usepackage{xcolor}
\def\BibTeX{{\rm B\kern-.05em{\sc i\kern-.025em b}\kern-.08em
    T\kern-.1667em\lower.7ex\hbox{E}\kern-.125emX}}
\usepackage{multirow}
\usepackage{booktabs}
\usepackage{amsfonts}
\usepackage{amsmath}
\usepackage{amsmath,cases}
\usepackage{amssymb}
\usepackage{blindtext, graphicx}
\usepackage{cite}
\usepackage{epsfig}
\usepackage{url}
\usepackage{algorithm}
\usepackage{algorithmicx, algpseudocode}
\usepackage{epstopdf}
\usepackage{color}
\usepackage[tight]{subfigure}
\usepackage{mathrsfs}
\usepackage[justification=centering]{caption}
\usepackage{float}
\usepackage{cuted}
\usepackage{lipsum}
\usepackage{diagbox}
\usepackage{graphicx,tabularx}

\newcommand{\ds}{\displaystyle}

\newcommand{\la}{\langle}
\newcommand{\ra}{\rangle}

\newcommand{\clC}{{\cal C}}

\newcommand{\clK}{{\cal K}}

\newcommand{\clR}{{\cal R}}

\newcommand{\bw}{\mathbf{w}}

\newcommand{\clB}{{\cal B}}

\newcommand{\clH}{{\cal H}}

\newcommand{\alphak}{\alpha^{(\kappa)}}

\newcommand{\rk}{r^{(\kappa)}}
\newcommand{\ak}{a^{(\kappa)}}

\newcommand{\btheta}{\pmb{\theta}}
\newcommand{\thetak}{\theta^{(\kappa)}}
\newcommand{\thetako}{\theta^{(\kappa+1)}}

\newcommand{\wko}{w^{(\kappa+1)}}

\newcommand{\wk}{w^{(\kappa)}}

\newcommand{\tak}{\tilde{a}^{(\kappa)}}

\newcommand{\bk}{b^{(\kappa)}}
\newcommand{\ck}{c^{(\kappa)}}
\newcommand{\tbk}{\tilde{b}^{(\kappa)}}

\newcommand{\tck}{\tilde{c}^{(\kappa)}}

\newcommand{\bbC}{\mathbb{C}}

\newcommand{\Psik}{\Psi^{(\kappa)}}
\newcommand{\hPsik}{\hat{\Psi}^{(\kappa)}}

\newcommand{\tdh}{\tilde{h}}

\newcommand{\cWko}{{\cal W}^{(\kappa+1)}}

\newcommand{\betak}{\beta^{(\kappa)}}
\newcommand{\rgk}{\gamma_k}
\newcommand{\rgj}{\gamma_j}
\newcommand{\rgkk}{\rgk^{(\kappa)}}
\newcommand{\rgjk}{\rgj^{(\kappa)}}

\newcommand{\gammak}{\gamma^{(\kappa)}}

\newcommand{\hr}{\hat{r}}

\newcommand{\hf}{\hat{f}}

\newcommand{\hfk}{\hf^{(\kappa)}}

\newcommand{\clHk}{\clH^{(\kappa)}}
\newcommand{\hrk}{\hat{r}^{(\kappa)}}

\newcommand{\thfk}{\tilde{\hat{f}}^{(\kappa)}}
\newcommand{\ellk}{\ell^{(\kappa)}}

\newcommand{\upsilonk}{\upsilon^{(\kappa)}}

\allowdisplaybreaks
\begin{document}
\title{\Large Finite-Blocklength RIS-Aided Transmit Beamforming}
\author{M. Abughalwa, H. D. Tuan, D. N. Nguyen, H. V. Poor, and
L. Hanzo
\thanks{The work of  M. Abughalwa was supported by an Australian
Government Research Training Program Scholarship. The work of H. D. Tuan was supported by the Australian Research Council's Discovery Projects under Grant DP190102501. The work of H. V. Poor was supported by the U.S. National Science Foundation under Grant CNS-2128448. The work of L. Hanzo was supported by
the Engineering and Physical Sciences Research Council projects EP/P034284/1 
and EP/P003990/1 (COALESCE), and  the European Research 
Council's Advanced Fellow Grant QuantCom under Grant 789028.}
\thanks{M. Abughalwa, H. D. Tuan, and  D. N. Nguyen are with the 
School of Electrical and Data Engineering, University of Technology Sydney,
NSW 2007, Australia (email: monir.abughalwa@student.uts.edu.au, tuan.hoang@uts.ed.au, diep.nguyen@uts.ed.au).}
\thanks{H. V. Poor is with the Department of Electrical and Computer Engineering, Princeton University, NJ 08544, USA (email:poor@princeton.edu).}
\thanks{L. Hanzo is with the School of Electronics and Computer Science, University of Southampton, Southampton, SO17 1BJ, UK (email: lh@ecs.soton.ac.uk).}
}
\maketitle
\begin{abstract}
This paper considers the downlink of an ultra-reliable low-latency communication (URLLC) system in which
a base station (BS) serves multiple single-antenna users in the  short (finite) blocklength (FBL)
regime  with the assistance of a reconfigurable intelligent surface (RIS). In the FBL regime, the users' achievable rates are  complex functions of  the beamforming vectors and  of the RIS's programmable reflecting elements (PREs).
We propose the joint design of the transmit beamformers and PREs, the problem of maximizing the geometric mean (GM) of  these rates (GM-rate) and show that this aforementioned results are providing fair rate distribution and thus reliable
links to all users.
A novel computational algorithm is developed, which is based on closed forms to generate improved feasible points, using its execution. The simulations show the merit of our solution.
\end{abstract}


\section{Introduction} \label{sec1}
Reconfigurable intelligent surfaces (RIS) constituted a planar array of passive programmable reflecting elements (PREs),
intentionally augment  the coverage of future wireless networks (6G)
~\cite{Reetal20,Huaetal21,CGZZ21}. Explicitly their spectral efficiency can be maximized
 by   the joint design of the transmit beamformer (TBF)
at the BS and the RIS PREs \cite{Hanetal19tvt,HMY20,Yuetal21,Yuetal21twc}.

Ultra-reliable and low-latency communication (URLLC) has also attracted recent research attention thanks to its potential applications in the internet of things (IoT), with special attention to, holographic communications, the tactile Internet, autonomous driving etc. ~\cite{BrownQualcomm2018,ben18}. Under the URLLC framework, low-latency requires short (finite) blocklength (FBL) while ultra-reliability imposes extra low error probability constraints  \cite{PPV10}. As a consequence, the rate function of URLLC is dependent not only on the signal-to-noise ratio (SNR) but also on the blocklength and the decoding error probability.
Hence, its definition is much more computationally challenging  than that of the  Shannon's rate function in the long block regime. Resource allocation and transmit beamforming used for optimizing the users' rate under the FBL regime have been recently considered e.g. in \cite{Nasetal21a,Nasetal21b}.

The authors of \cite{Hasetal21,RWP21} analysed RIS-aided URLLC systems of a single antenna BS and a RIS serving a single user. The more advanced joint design of the transmit beamformer
at multiple BSs and RIS PREs maximizing  the sum-rate subject to specific quality of service (QoS) constraints in terms of the
users' rates was considered in  \cite{GJS21}. However, the computational complexity of the algorithm proposed in
\cite{GJS21} is extremely high, as it iterates by observing convex problems of escalating dimension.
Hence Ghanem et al. \cite{GJS21}
considered  only up to $20$ PREs for the RIS, even though  RIS should employ very large numbers of PREs  \cite{BOL19}. {\color{black}Regarding this problem, one can combine the techniques proposed in \cite{Yuetal21} and  \cite{Nasetal21a}
to develop an algorithm, which iterates by evaluating convex problems of the same size as the original nonconvex problem.
However, this size is already large  for practical RIS-aided networks
due to the large numbers of  PREs and beamforming decision variables, which makes the
computation of these convex problems not really tractable.}

Against the above background, this paper provides a computationally tractable solution for the joint design of TBF and RIS PREs to optimize all users's rate in the FBL regime. Following our earlier results in   \cite{Yuetal21twc} for optimizing all users' rates in the long blocklength (LBL) regime (Shannon rate), we now aim for maximizing the geometric means of the users' rates (GM-rate) as we explicitly demonstrate it is capable of providing a fair users' rate distribution without
enforcing computationally intractable rate constraint. As a further novelty, we avoid the computationally intractable unit modulus constraints on the PREs by directly optimizing their argument. As such, the design of PREs is based on trigonometric function optimization.

The rest of the paper is organized as follows Section II is devoted to the problem statement and solution, which is
supported by our simulation results provided in Section III. Finally, Section IV concludes the paper.

{\it Notation.} Only the vector variables are printed in boldface; $I_N$ is the identity matrix of size $N\times N$;
For $x=(x_1,\dots, x_n)^T$, ${\sf diag}(x)$ is a diagonal matrix of the size $n\times n$ with $x_1, x_2, \dots, x_n$ on its diagonal; $\la x,y\ra=x^Hy$ is the dot product of
the vectors $x$ and $y$; The notation $X \succeq 0$ ($X\succ 0$, resp.) used for the Hermitian symmetric matrix $X$ indicates that it is positive semi-definite
(positive definite, resp.); The maximal eigenvalue of the Hermitian symmetric matrix $X$ is denoted by $\lambda_{\max}(X)$;  For a real-valued vector $x=(x_1,\dots, x_n)^T\in\mathbb{R}^n$, $e^{\jmath x}$ is entry-wise understood,
i.e.  $e^{\jmath x}=(e^{\jmath x_1}, \dots, e^{\jmath x_n})^T\in\mathbb{C}^n$.
$\angle x\in [0,2\pi)$  is the argument of the complex $x$.
\section{Problem statement} \label{sec2}
\begin{figure}[!htb]
	\centering
	\includegraphics[width=0.4\textwidth]{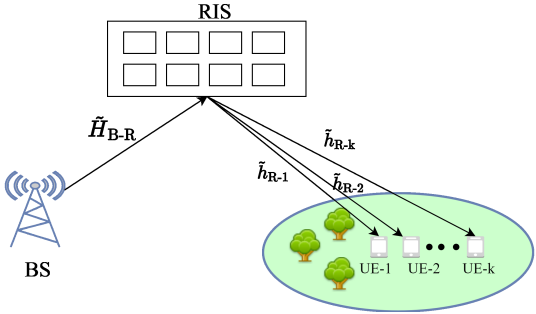}
	\caption{System model}
	\label{s1}
\end{figure}
As illustrated by Fig. \ref{s1}, we consider the downlink of a system, in which  an $M$-antenna BS
serves $K$ single-antenna users (UEs) $k\in\clK\triangleq \{1,\dots, K\}$ with the aid of a RIS having
$N$ PREs as the BS cannot see the UEs. As the RIS is seen by the BS and the UEs are seen by the RIS
line-of-sight (LoS),
the channels spanning from the BS to the RIS (from the RIS to UE $k$, resp.) are modelled by $\tilde{H}_\text{B-R}=\sqrt{\beta_\text{B-R}}H_\text{B-R}\in\mathbb{C}^{N\times M}$
($\tilde{h}_\text{R-k}=\sqrt{\beta_\text{R-k}} h_\text{R-k}\in\mathbb{C}^{1\times N}$, resp.), where
$\sqrt{\beta_\text{R-k}}$ and  $\sqrt{\beta_\text{B-R}}$  respectively represents the
path-loss and large-scale fading of  the RIS-to-UE $k$ link and the BS-to-RIS link, while $h_\text{R-k}$ and $H_\text{B-R}$ are modelled by Rician fading \cite{Naetal19}.

Let $s_k\in \clC(0,1)$ be the information symbol intended for UE $k$, which is beamformed by the array of weights  $\bw_{k}\triangleq (\bw_k(1),\dots, \bw_k(M))^T \in\mathbb{C}^M$ to create the transmit signal $x=\sum_{k\in\clK}\bw_{k}s_k$.
The signal received at UE $k$ is given by
\begin{eqnarray}
y_k=\clH_k(\btheta)\sum_{k\in\clK}\bw_{k}s_k+n_k,\label{ps2}
\end{eqnarray}
for $\clH_k(\btheta)\triangleq\tdh_\text{BR-k}{\sf diag}(e^{\jmath \btheta}) H_\text{B-R}=\sum_{n=1}^N\clH_{k,n}e^{\jmath \btheta_n}\in\bbC^{1\times M}$,
with $\tdh_\text{BR-k}\triangleq \sqrt{\beta_\text{B-R}}   \sqrt{\beta_\text{R-k}} h_\text{R-k} \clR^{1/2}_\text{R-k}\in
\mathbb{C}^{1\times N}$,
where $\clR_\text{R-k} \in \bbC^{N\times N}$ encompasses the spatial correlation of the PREs with respect to  user $k$ \cite{Naetal19}, $n_k\in \clC(0,\sigma)$ is the background noise, and
 $\btheta=(\btheta_1,\dots, \btheta_N)^T\in [0,2\pi)^N$
is the vector of the PREs' angles, and $\clH_{k,n}\triangleq \tdh_\text{BR-k}\Upsilon_n H_\text{B-R}$, where
$\Upsilon_n$ is the matrix of size $N\times N$ with all-zero entries apart from $\Upsilon_n(n,n)=1$.

For $\bw\triangleq \{\bw_k, k\in\clK\}$, the effective signal-to-interference-plus-noise (SINR) at UE $k$ is defined by
\begin{equation}\label{sinr1}
g_k(\bw, \btheta)=\frac{|\clH_k(\btheta)\bw_k|^2}{\alpha_k(\bw,\btheta)},
\end{equation}
for $\alpha_k(\bw,\btheta)\triangleq \sum_{j\in\clK\setminus\{k\}}|\clH_k(\btheta)\bw_j|^2+\sigma$.
In the LBL regime, the rate in {\color{black}nats/sec/Hz} at UE $k$ is
$r_k(\bw,\btheta)=\ln\left[1+ g_k(\bw, \btheta)\right]$.

Let $\clB$ be the communication bandwidth.  According to \cite{STD17}, by treating other terms there as Gaussian noise the achievable URLLC rate in nats/sec/Hz
for the signal $s_k$ in \eqref{ps2} is approximated by
\begin{equation}\label{snr5}
\hr_k(\bw,\btheta) \triangleq  r_k(\bw,\btheta)- a\upsilon^{1/2}_k(\bw,\btheta),
\end{equation}
where the channel included dispersion $\upsilon_k(\bw,\btheta)$ under the SINR $g_k(\bw, \btheta)$ is defined by
\cite[eq. (27)]{STD17}
\begin{eqnarray}
\upsilon_k(\bw,\btheta)&\triangleq& 2  \frac{g_k(\bw, \btheta)}{1+g_k(\bw, \btheta)}\label{snr6}\\
&=&2\left(1-\frac{\alpha_k(\bw,\btheta)}{\beta_k(\bw,\btheta)}\right),
\end{eqnarray}
in conjunction with
\begin{equation}\label{beta}
\beta_k(\bw,\btheta)\triangleq \alpha_k(\bw,\btheta)+|\clH_k(\btheta)\bw_k|^2= \sum_{j\in\clK}|\clH_k(\btheta)\bw_j|^2+\sigma.
\end{equation}
Also, $a\triangleq \frac{1}{\sqrt{\clB t_t}}Q^{-1}_G(\epsilon^c)$,
where $t_t$ is the URLLC transmission duration, $Q^{-1}_G(.)$ is the inverse of the Gaussian Q-function $Q(x)=\int_x^{\infty}\frac{1}{\sqrt{2\pi}}\exp(-t^2/2)dt$, and {\color{black}$\epsilon^c$ is defined as an acceptable decoding error probability, which implies that under the block fading channel model  considered, one out of $1/\epsilon_c$ short packets (URLLC transmissions) may experience outage.
	
We consider the following problem of jointly designing the beamformer $\bw$ and the PREs $\btheta$
for maximizing the GM-rate:
\begin{subequations}\label{aps4r}
	\begin{eqnarray}
	\max_{\bw,\btheta}\hf(\bw,\btheta)\triangleq \left(\prod_{k=1}^{K} \hr_k(\bw,\btheta)\right)^{1/K}\label{aps4ra}\\ \mbox{s.t.}\quad \sum_{k=1}^K||\bw_{k}||^2\leq P,\label{ps4b}
	\end{eqnarray}
\end{subequations}
where (\ref{ps4b}) sets the transmit sum power constraint within a  given power budget $P$. Our previous paper \cite{Yuetal21twc}, which considered the problem
\begin{equation}\label{lbl1}
\max_{\bw,\btheta}f(\bw,\btheta)\triangleq \left(\prod_{k=1}^{K} r_k(\bw,\btheta)\right)^{1/K}\quad\mbox{s.t.}\quad
(\ref{ps4b})
\end{equation}
in the LBL regime showed that GM-rate maximization naturally leads to fair user rate distributions without imposing the rate constraints of  $r_k(\bw,\btheta)\geq \bar{r}$, which are nonconvex and thus computationally intractable.  As a compelling benefit, by directly optimizing the angles $\btheta\in [0,2\pi)^{N}$ of
PREs, both (\ref{aps4r}) and (\ref{lbl1}) circumvent  the unit modulus constraints on the latters.

Compared to the rate function $r_k$ of the LBL regime, we have the rate-reduction term $\upsilon^{1/2}_k(\bw,\btheta)$ arisen in the rate function $\hr_k$ in the FBL regime. Therefore, the main challenge in considering (\ref{aps4r}) is to handle this term.

Initialized by $(w^{(0)}, \theta^{(0)})$ as the optimal solution of (\ref{lbl1}) computed by \cite{Yuetal21twc},
let $(\wk,\thetak)$ be a feasible point for (\ref{aps4r}) that is found from the
$(\kappa-1)$-st round, and
\begin{equation}\label{gk2}
	\gammak_k\triangleq \frac{\max_{k'\in\clK}\hr_{k'}(\wk,\thetak)}{\hr_k(\wk,\thetak)}, k\in\clK.
\end{equation}
As discussed in \cite{Yuetal21twc}, the descent iterations are based on the following problem
\begin{equation}\label{conin4}
\max_{\bw,\btheta}\hfk(\bw,\btheta)\triangleq \sum_{k=1}^K\gammak_k \hr_k(\bw,\btheta)\quad\mbox{s.t.}\quad (\ref{ps4b}).
\end{equation}
\subsection{Beamforming descent iteration}
We seek $w^{(\kappa+1)}$ satisfying:
\begin{equation}\label{t1}
\hfk(\wko,\thetak)>\hfk(\wk,\thetak).
\end{equation}
Define $\rk_{b,k}(\bw)\triangleq r_k(\bw,\thetak), \upsilonk_{b,k}(\bw)\triangleq \upsilon_k(\bw,\thetak)$, $\hrk_{b,k}(\bw)\triangleq \hr_k(\bw,\thetak)$, $\alphak_{b,k}(\bw)\triangleq \alpha_k(\bw,\thetak)$,
$\betak_{b,k}(\bw)\triangleq \beta_k(\bw,\thetak)$, $\upsilonk_{b,k}(\bw)\triangleq \upsilon_k(\bw,\thetak)$, and $\clHk_k\triangleq \clH_k(\thetak)$. As such,
\begin{equation}\label{t1b}
\hfk(\bw,\thetak)=\sum_{k=1}^K\gammak_k \hrk_{b,k}(\bw),
\end{equation}
and
\begin{equation}\label{t1a}
\hrk_{b,k}(\bw)=\rk_{b,k}(\bw)-a\sqrt{\upsilonk_{b,k}(\bw)}.
\end{equation}
The following lower bounding
concave approximation of $\rk_{b,k}(\bw)$   was obtained
in \cite{Yuetal21}:
\begin{eqnarray}
\rk_{b,k}(\bw)\geq\ak_{k,1}+ 2\Re\{\la \bk_{k,k},\bw_k\ra\}-\ck_{k,1}\sum_{j=1}^K|\clHk_k\bw_j|^2,\label{bo3a}
\end{eqnarray}
with $\ak_{k,1}\triangleq \rk_{b,k}(\wk)-g_k(\thetak,\wk)-\sigma\ck_{k,1}$,
$\bk_{k,k}\triangleq (\clHk_k)^H\clHk_k\wk_k/\alphak_{b,k}(\wk)$,
$0< \ck_{k,1}\triangleq 1/\alphak_{b,k}(\wk)-1/\betak_{b,k}(\wk)$.

Our next step is now to develop an upper bounding convex approximation of $\sqrt{\upsilonk_{b,k}(\bw)}$,
which together with (\ref{bo3a}) gives a lower bounding concave approximation of  $\hrk_{b,k}$ in (\ref{t1a}).

Using the inequality
\begin{equation}\label{xy}
\sqrt{x}\leq \frac{\sqrt{\bar{x}}}{2}\left(1+\frac{x}{\bar{x}}\right)\quad\forall\ x>0, \bar{x}>0,
\end{equation}
gives
\begin{eqnarray}
\sqrt{\upsilonk_{b,k}(\bw)}&\leq&\frac{\sqrt{\upsilonk_{b,k}(\wk)}}{2}\left(1+\frac{2}{\upsilonk_{b,k}(\wk)}\right)
\nonumber \\ &&
-\frac{1}{\sqrt{\upsilonk_{b,k}(\wk)}}
\frac{\alphak_{b,k}(\bw)}{\betak_{b,k}(\bw)}. \label{xy2}
\end{eqnarray}
Applying the following inequality for $x\in\mathbb{C}^n$, $\bar{x}\in\mathbb{C}^n$, $y>0$, $\bar{y}>0$, and
$\sigma>0$,
\begin{eqnarray}
\frac{||x||^2}{y+\sigma}\geq \frac{||\bar{x}||^2}{\bar{y}+\sigma}\left(2\frac{\Re\{\bar{x}^Hx\}}{||\bar{x}||^2}
-\frac{y+\sigma}{\bar{y}+\sigma} \right) \label{xt}
\end{eqnarray}
 yields:
\begin{align*}
	&\frac{\alphak_{b,k}(\bw)}{\betak_{b,k}(\bw)}
	\geq \frac{\alphak_{b,k}(\wk)}{\betak_{b,k}(\wk)} \\
	&\times \left(2\frac{\sum_{j\in\clK\setminus\{k\}}\Re\{(\wk_j)^H[(\clHk_k)^H]^2\bw_j\}+\sigma }{\alphak_{b,k}(\wk)} - \right. \nonumber \\
	& \left. \frac{\betak_{b,k}(\bw)}{\betak_{b,k}(\wk)} \right),
\end{align*}
which together with (\ref{xy2}) gives the following upper bounding convex approximation of $a\sqrt{\upsilonk_{b,k}(\bw)}$:
\begin{eqnarray}
a\sqrt{\upsilonk_{b,k}(\bw)}&\leq& \ak_{k,2}-2\sum_{j\in\clK\setminus\{k\}}\Re\{\la\bk_{k,j},\bw_j \ra  \}+
\nonumber \\
&& \ck_{k,2}\sum_{j=1}^K|\clHk_k\bw_j|^2,\label{v12}
\end{eqnarray}
for $\ak_{k,2}\triangleq\ds  a\frac{\sqrt{\upsilonk_{b,k}(\wk)}}{2}\left(1+\frac{2}{\upsilonk_{b,k}(\wk)}\right)
+  \sigma\frac{\alphak_{b,k}(\wk)}{\betak_{b,k}(\wk)\sqrt{\upsilonk_{b,k}(\wk)}}
\left(\frac{-2}{\alphak_{b,k}(\wk)}+\frac{1}{\betak_{b,k}(\wk)} \right)$,
$\bk_{k,j}\triangleq \frac{a}{\betak_{b,k}(\wk)\sqrt{\upsilonk_{b,k}(\wk)}}(\clHk_k)^H\clHk_k\wk_j$,
$j\in\clK\setminus\{k\}$,
and
$\ck_{k,2}\triangleq a\frac{\alphak_{b,k}(\wk)}{(\betak_{b,k}(\wk))^2\sqrt{\upsilonk_{b,k}(\wk)}}$.

The bounds (\ref{bo3a}) and (\ref{v12}) yield the following lower bounding concave approximation for
$\hrk_{b,k}(\bw)$ in (\ref{t1a}):
\begin{eqnarray}
\hrk_{b,k}(\bw) \geq  \ak_k+2\sum_{j=1}^K\Re\{\la\bk_{k,j},\bw_j \ra-
\ck_k\sum_{j\in\clK}|\clHk_k\bw_j|^2,\label{rhk}
\end{eqnarray}
for $\ak_k\triangleq \ak_{k,1}-\ak_{k,2}$, and $\ck_k\triangleq \ck_{k,1}+\ck_{k,2}$.

We now generate $\wko$ as the optimal solution of the following problem
\begin{equation}\label{pbo2}
\max_{\bw}\ \hfk_b(\bw)\quad\mbox{s.t.}\quad (\ref{ps4b}),
\end{equation}
where $\hfk_b(\bw)\triangleq\sum_{k=1}^{K}\gammak_k\ak_k+2\sum_{k=1}^{K}\Re\{\la \bk_k,\bw_k\ra\}-  \sum_{k=1}^{K}(\bw_k)^H\Psik_b\bw_k$
with $\bk_k\triangleq \sum_{j=1}^K\gammak_j\bk_{j,k}$,
and $0\preceq \Psik_b\triangleq \sum_{j=1}^K\rgjk\ck_j(\clHk_j)^H\clHk_j$. It can be readily checked that
\begin{equation}\label{check1}
\hfk(\wk,\thetak)=\hfk_b(\wk).
\end{equation}
The problem (\ref{pbo2})
admits the following closed-form solution
\begin{equation}\label{sr3}
\wko_k=\begin{cases}\begin{array}{l}(\Psik_b)^{-1}\bk_k\quad \mbox{if}\quad \ds\sum_{k=1}^{K}
||(\Psik_b)^{-1}\bk_k||^2\leq P\\
\left(\Psik_b+\mu I_M \right)^{-1}\bk_k\quad \mbox{otherwise},
\end{array}
\end{cases}
\end{equation}
where $\mu>0$ is chosen by bisection such that $\sum_{k=1}^{K}||\left(\Psik_b+\mu I_M\right)^{-1}\bk_k||^2= P$.

It follows from (\ref{t1b}) and (\ref{rhk}) that $\hfk(\wko,\thetak)\geq  \hfk_b(\wko)$, while
$\hfk_b(\wko)>\hfk_b(\wk)=\hfk(\wk,\thetak)$, because $\wko$ and $\wk$ represent the optimal solution and a feasible point for (\ref{pbo2}). We thus have (\ref{t1}) as sought.
\subsection{Programmable reflecting elements' descent iteration}
We seek the next iterative point $\theta^{(\kappa+1)}$ such that
\begin{equation}\label{pp0}
\hfk(\wko,\thetako)>\hfk(\wko,\thetak).
\end{equation}
Define $\rk_{p,k}(\btheta)\triangleq r_k(\wko,\btheta)$,
$\upsilonk_{p,k}(\btheta)\triangleq \upsilon_k(\wko,\btheta)$,
$\hrk_{p,k}(\btheta)\triangleq \hr_k(\wko,\btheta)$,
$\alphak_{p,k}(\btheta)\triangleq \alpha_{k}(\wko,\btheta)$,
$\betak_{p,k}(\btheta)\triangleq \beta_k(\wko,\btheta)$,
$\upsilonk_{p,k}(\btheta)\triangleq \upsilon_k(\wko,\btheta)$, and
$\ellk_{k,j}\triangleq \clH_k(\thetak)\wko_j$, $(k,j)\in\clK\times\clK$. As such,
\begin{equation}\label{p1b}
\hfk(\wko,\btheta)=\sum_{k=1}^K\gammak_k \hrk_{p,k}(\btheta),
\end{equation}
and
\begin{equation}\label{p1a}
\hrk_{p,k}(\btheta)=\rk_{p,k}(\btheta)-a\sqrt{\upsilonk_{p,k}(\btheta)}.
\end{equation}
Recall that $\clH_{k,n}$ are defined in (\ref{ps2}).
The following lower bounding
 approximation of $\rk_{p,k}(\btheta)$  was obtained
in \cite{Yuetal21}:
\begin{eqnarray}
\rk_{p,k}(\btheta)&\geq& \tak_{k,1}+ 2\Re\{\sum_{n=1}^N\tbk_{k,1}(n)e^{\jmath \btheta_n}\}
- \nonumber \\
&&\tck_{k,1}(e^{\jmath\btheta})^H\Psik_p e^{\jmath\btheta},\label{pbo7}
\end{eqnarray}
with $\tak_{k,1}\triangleq \rk_{p,k}(\thetak)-g_k(\thetak,\wko)-\sigma \tck_{k,1}$,
$0< \tck_{k,1}\triangleq 1/\alphak_{p,k}(\thetak)-1/\betak_{p,k}(\thetak)$,
and
$\tbk_{k,1}(n)\triangleq \ds\frac{1}{\alphak_{p,k}(\thetak)} (\ellk_{k,k})^H\clH_{k,n}\wko_k$,
$n=1,\dots N$,
$\cWko\triangleq \sum_{j=1}^{K}[\wko_j]^2$,
$\Psik_p(n',n)\triangleq  \la \clH_{k,n}\cWko\clH^H_{k,n'}\ra$, $n'=1,\dots, N$;
$n=1,\dots, N$.

Similarly to (\ref{v12}), we have
\begin{eqnarray}
a\sqrt{\upsilonk_{p,k}(\btheta)}&\leq& \tak_{k,2}-2\Re\{\sum_{n=1}^N\tbk_{k,2}(n)e^{\jmath\btheta_n}\}
+ \nonumber \\
&& \tck_{k,2}\sum_{j=1}^K(e^{\jmath \btheta})^H\Psik_p e^{\jmath \btheta},\label{av12}
\end{eqnarray}
for $\tak_{k,2}\triangleq a\frac{\sqrt{\upsilonk_{p,k}(\thetak)}}{2}\left(1+\frac{2}{\upsilonk_{p,k}(\thetak)}\right)
+  \sigma\frac{\alphak_{p,k}(\thetak)}{\betak_{p,k}(\thetak)\sqrt{\upsilonk_{p,k}(\thetak)}}
\left(\frac{-2}{\alphak_{p,k}(\thetak)}+\frac{1}{\betak_{p,k}(\thetak)} \right)$,
and
$\tbk_{k,2}(n)\triangleq \frac{a}{\betak_{p,k}(\thetak)\sqrt{\upsilonk_{p,k}(\thetak)}} \sum_{j\in\clK\setminus\{k\}}
(\ellk_{k,j})^H\clH_{k,n}\wko_j$,
$n=1,\dots, N$,
and
$\tck_{k,2}\triangleq a\frac{\alphak_{p,k}(\thetak)}{(\betak_{p,k}(\thetak))^2\sqrt{\upsilonk_{p,k}(\thetak)}}$.
Based on (\ref{pbo7}) and (\ref{av12}) we obtain the following lower bound:
\begin{eqnarray}
\hrk_{p,k}(\btheta)\geq\tak_{k}+2\Re\{\sum_{n=1}^N\tbk_{k}(n)e^{\jmath \btheta_n}\}-
\tck_k(e^{\jmath \btheta})^H\Psik_p e^{\jmath \btheta},
\end{eqnarray}
where $\tak_k\triangleq \tak_{k,1}-\tak_{k,2}$, $\tck_k\triangleq \tck_{k,1}+\tck_{k,2})$,
and $\tbk_k\triangleq \tbk_{k,1}+\tbk_{k,2}$. Then,
\begin{eqnarray}
\hfk(\wko,\btheta)&\geq& \tak+2\Re\{\sum_{n=1}^N\tbk(n)e^{\jmath \btheta_n}\} \nonumber \\
&& -(e^{\jmath \btheta})^H\hPsik_p e^{\jmath \btheta},\label{pbo12}
\end{eqnarray}
for $\tak\triangleq \ds\sum_{k=1}^K\rgkk\tak_k$,
$\tbk(n)\triangleq \ds\sum_{k=1}^K\rgkk\tbk_k(n), n=1,\dots, N$, and
$
0\preceq \hPsik_p\triangleq \left(\sum_{k=1}^K\gammak_k\tck_k\right)\Psik_p$.
Furthermore, we have
\begin{equation}\label{ppo3}
\mbox{RHS of (\ref{pbo12})}\geq \hfk_p(\btheta)
\end{equation}
for  $\hfk_p(\btheta)\triangleq \tak+2\Re\{\sum_{n=1}^N(\tbk(n)-\sum_{m=1}^N e^{-\jmath \thetak_m}
\hPsik_p(m,n)
+\lambda_{\max}(\hPsik_p)e^{-\jmath \thetak_n})e^{\jmath\btheta_n}\}
-(e^{\jmath \thetak})^H\hPsik_p e^{\jmath \thetak}-2\lambda_{\max}(\hPsik_p)N$.
We thus generate $\thetako$ as the optimal solution of the problem
\begin{equation}\label{ppo4}
\max_{\btheta}\ \thfk_p(\btheta),
\end{equation}
which admits the closed-form solution\footnote{$[(\hPsik_p-\mu I_N)e^{\jmath \thetak}](n)$ is the $n$-th
	entry of $(\hPsik_p-\mu I_N)e^{\jmath \thetak}$} of
\begin{align}\label{ppo5}
\thetako_n =& 2\pi-\angle\left(\tbk(n)-\sum_{m=1}^N e^{-\jmath \thetak_m}
\hPsik_p(m,n)+ \right. \nonumber \\
& \left. \lambda_{\max}(\hPsik_p)e^{-\jmath \thetak_n}\right), n=1,\dots, N.
\end{align}
It follows from (\ref{ppo3}) that $\hfk(\wko,\thetako)\geq \hfk_p(\thetako)
\geq \thfk_p(\thetako)> \thfk_p(\thetak)=\hfk_p(\thetak)
=\hfk(\wko,\thetak)$, confirming (\ref{pp0}). Hence $\thetako$ is a better feasible point than $\thetak$.
\subsection{Algorithm}
Algorithm \ref{alg1} provides the pseudo-code for the proposed steep descent computational procedure of (\ref{aps4r}) as the iterations (\ref{sr3}) and (\ref{ppo5}) seek a descent direction by seeking a better feasible point for the nonconvex problem (\ref{aps4r}).

\begin{algorithm}
	\caption{URLLC GM-rate descent algorithm} \label{alg1}
	\begin{algorithmic}[1]
		\State \textbf{Initialization:}  Use the Algorithm \cite{Yuetal21twc} to initialize
		a feasible $(w^{(0)},\theta^{(0)})$. Set $\kappa=0$.
		\State \textbf{$\kappa$-th iteration:} Generate $\wko$ by (\ref{sr3}) and
		$\thetako$ by (\ref{ppo5}). {\color{black} Given the convergence tolerance $\nu_t$, stop
if $|\hat{f}(\wko,\thetako)-\hat{f}(\wk,\thetak)|/\hat{f}(\wk,\thetak)\leq \nu_t$.}
Reset $\kappa\leftarrow \kappa+1$.
		\State \textbf{Output} $(\wk, \thetak)$ and URRLC rates $\hr_k(\wk,\thetak)$, $k\in\clK$ with their
		GM $\left( \prod_{k=1}^K\hr_k(\wk,\thetak)\right)^{1/K}$.
	\end{algorithmic}
\end{algorithm}

{\color{black}Remark. One can see that the above Algorithm
invokes the nonconvex problem (\ref{conin4}) at each iteration, which is a problem of weighted sum rate maximization associated with the weights
$\gammak_k$ iteratively updated according to (\ref{gk2}), to generate a better feasible point.
}
\section{Numerical examples} \label{sec3}
This section evaluates the performance of the proposed algorithm using numerical examples.
The set up is the same as that in \cite{Yuetal21twc}: the large-scale fading and the RIS-to-UE $k$ path-loss is ${\beta_\text{R-k}} = G_{\text{RIS}}-33.05-30 \log_{10}(d_{\text{R-k}})$ dB, where $d_{\text{R-k}}$ is the distance between the RIS and UE $k$ in meters, while $G_{\text{IRS}}$ is the antenna gain of the RIS elements \cite{Naetal19,BOL19}. The large-scale fading and the path-loss between BS and RIS is  ${\beta_\text{B-R}} = G_{\text{\text{BS}}}+G_{\text{IRS}}-35.9-22 \log_{10}(d_{\text{B-R}})$ dB, where $d_{\text{B-R}}$ is the distance between the BS and RIS in meters, while $G_{\text{\text{BS}}}$ is the BS antenna gain \cite{Naetal19,BOL19}. The
coordinates of the BS and the RIS are $(20,0,25)$ and  $(0,30,40)$, while
the users are randomly located  in a $(60m \times 60m)$ area to the right of the BS and RIS. The entries
of the BS-to-RIS LoS channel matrix are $[H_{B-R}]_{n,m} = e^{j\pi \left((n-1) \sin\overline{\theta}_n \sin \overline{\phi}_n+(m-1) \sin\theta_n \sin\phi_n\right)}$, where ${\theta}_n$ and ${\phi}_n$ are uniformly distributed over $(0,\pi)$ and $(0,2 \pi)$, and $\overline{\theta}_n = \pi-\theta_n$, $\overline{\phi}_n = \pi+\phi_n$. The small-scale fading channel gain $h_{\text{R-k}}$ follows the Rician distribution having K-factor of $3$. The spatial correlation matrix is $[R_{R-k}]_{n,n'} = e^{j\pi (n-n') \sin \tilde{\phi} \sin \tilde{\theta}}$, where $\tilde{\phi}$ and $\tilde{\theta}$ are the azimuth and elevation angle for UE $k$, respectively. Unless stated otherwise, the following parameters have been used in our simulation, $G_{\text{BS}}=G_{\text{IRS}} = 5$ dBi, $\clB = 1$ Mhz, $\sigma^2 = -174$ dBm/Hz, $M=10$, $K=10$, $P=20$ dBm, $N=100$, $\epsilon^c = 10^{-5}$, and $t_t=0.1$ ms which is suitable for URLLC transmission \cite{ben18}, and the choice of $1$ ms end-to-end delay ensures having a quasi-static channel during URLLC communication \cite{sh17cross}. The results are multiplied by $\log_2(e)$ to convert the unit nats/sec into the unit bps/Hz. {\color{black}Lastly we set the convergence tolerance $\nu_t$ to $10^{-3}$}.

Furthermore, we use the following terms for interpreting the results:
\begin{itemize}
	\item LBR refers to the performance in  LBL regime \cite{Yuetal21twc}.
	\item URLLC refers to the performance by Algorithm  \ref{alg1}.
\end{itemize}

\textcolor{black}{The problem of maximizing the sum rate (SR) can  also be solved by Algorithm \ref{alg1} upon
setting $\gammak_k \equiv 1$ in (\ref{gk2}). Note that the maximization of the SR represents the
maximization of the arithmetic mean (AM) of the users' rates (AM-rate)
because the latter is defined as the former divided by $K$.
Fig. \ref{AM-Fig1} plots the AM-rate achieved by maximizing
the GM-rate and SR. As expected,  the SR maximization achieves better AM-rate than GM-rate maximization.
However, SR maximization is unable to provide fairness for all users, as it assigns some users having low channel quality zero rate, which can be seen in Fig. \ref{PGS-Fig3} that plots the ratio between the minimum user rate and the maximum user rate (RR). As it can be seen, the RR under SR maximization is always zero because SR maximization
cannot avoid having zero rate in either LBR or in URLLC. This remains the case even
when the number $M$ of transmit antennas is higher than the number  $K$ of users. By contrast,
GM maximization manages to assign nonzero rates to all users, even when $M$ is lower than $K$. This
demonstrates that using GM rate maximization is capable of improving all users' rates.
Furthermore, Fig. 4 portrays the users' rate variance (URV) versus $M$ achieved by
GM-rate maximization and SR maximization. As expected the URV attained by SR-maximization is very high, as it tends to assign high fraction of the total SR to a few users. By contrast, the URV of GM-rate maximization is low and in fact it is not sensitive to the number $M$ of transmit antennas.}

Fig. \ref{PGS-Fig2} plots the GM rate versus $M$. As expected, the GM rate increases with $M$, since the system's ability to mitigate the multi user interference improving with $M$, especially when $M$ is higher than or equal to $K$. The URLLC GM rate increases similarly to the LBR GM rate. However, the gap between the LBR-GM rate and the URLLC-GM rate does not decrease with $M$ increasing.

\begin{figure}[t!]
	\centering
	\includegraphics[width=0.45\textwidth]{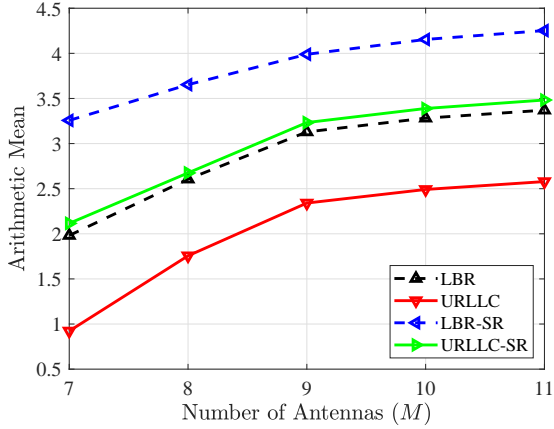}
	\caption{AM-rate  versus $M$}
	\label{AM-Fig1}
\end{figure}
\begin{figure}[t!]
	\centering
	\includegraphics[width=0.45\textwidth]{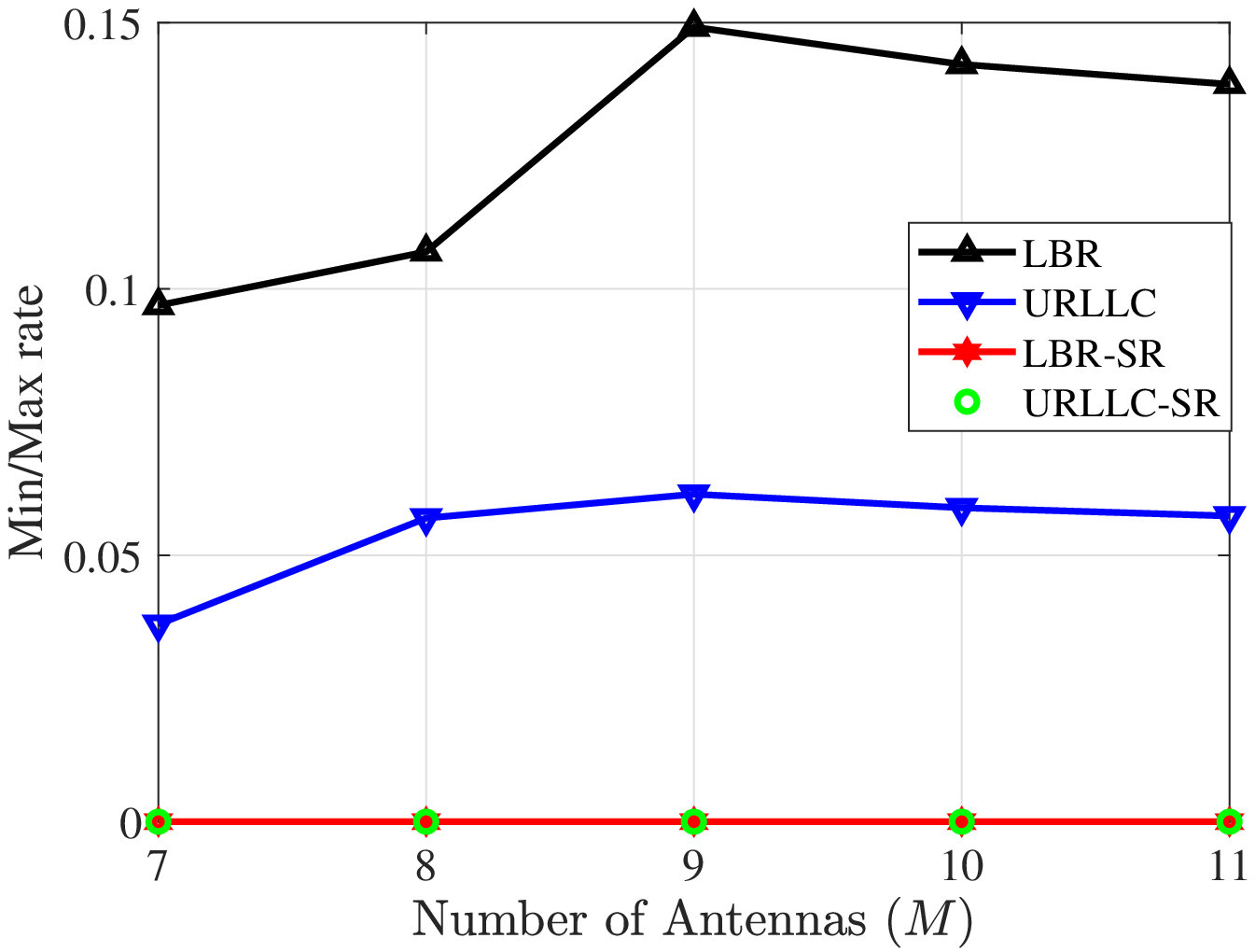}
	\caption{The ratio of the highest and lowest users rates versus $M$}
	\label{PGS-Fig3}
\end{figure}

\begin{figure}[t!]
	\centering
	\includegraphics[width=0.45\textwidth]{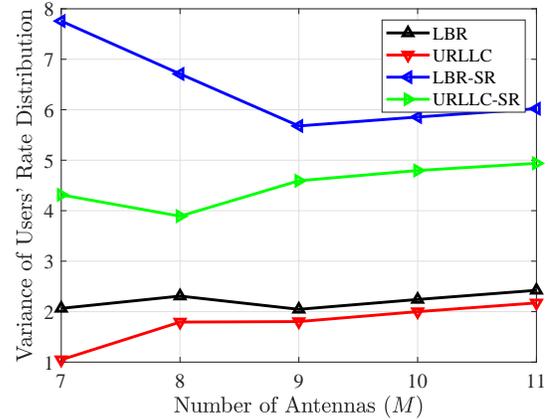}
	\caption{Users' Rate Variance versus $M$}
	\label{PGS-Figvar}
\end{figure}

\begin{figure}[t!]
	\centering
	\includegraphics[width=0.45\textwidth]{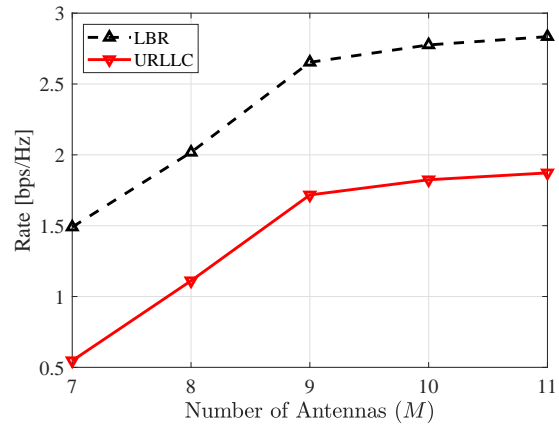}
	\caption{GM Rate versus $M$}
	\label{PGS-Fig2}
\end{figure}

\begin{figure}[t!]
	\centering
	\includegraphics[width=0.45\textwidth]{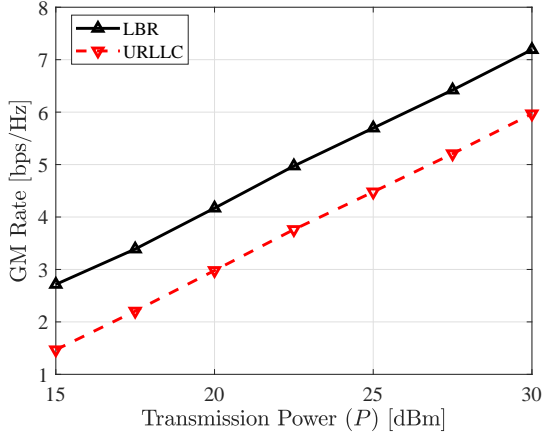}
	\caption{GM rate versus $P$}
	\label{PGS-Fig6}
\end{figure}

Fig. \ref{PGS-Fig6} depicts the GM rate of LBR and URLLC versus $P$, observe that as anticipated the GM rate increases with $P$. Furthermore, the URLLC behaviour is similar to LBR with the gap between the rates being almost the same, which indicates that increasing $P$ does not affect the overall URLLC rate.
\begin{figure}[t!]
	\centering
	\includegraphics[width=0.45\textwidth]{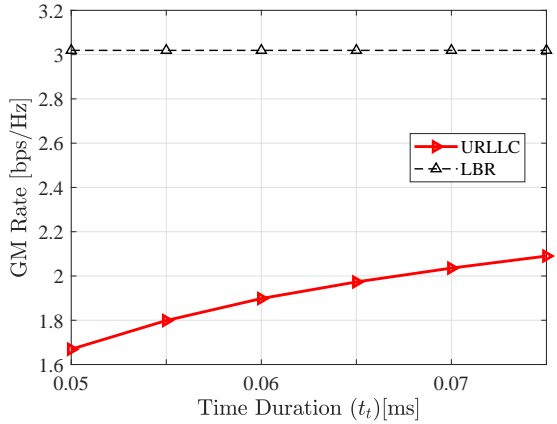}
	\caption{GM rate versus $t_t$}
	\label{PGS-Fig9}
\end{figure}

Fig. \ref{PGS-Fig9} portrays the URLLC-GM rate against $t_t$, which increases with $t_t$ but the rate increase gradually slows down. Nevertheless, even for low $t_t$ the system is still able to achieve a good rate, which represents the advantages of our Algorithm. \textcolor{black}{Note that the LBR assumes the transmission duration to be $\infty$, therefore it serve as an upper bound for the URLLC}. Moreover Fig. \ref{PGS-Fig9} can be used for choosing $t_t$ depending on the quality of service required.

\textcolor{black}{Lastly, Fig. \ref{PGS-Figtime} plots the RR versus $t_t$. The SR maximization in FBL  cannot avoid having zero rate even for long transmission duration of $t_t=0.1$ ms, while
GM maximization always assigns fair rates to all the users, regardless of the transmission duration $t_t$. Hence the advantage of using GM maximization over SR maximization becomes quite convincing.}

\begin{figure}[t!]
	\centering
	\includegraphics[width=0.45\textwidth]{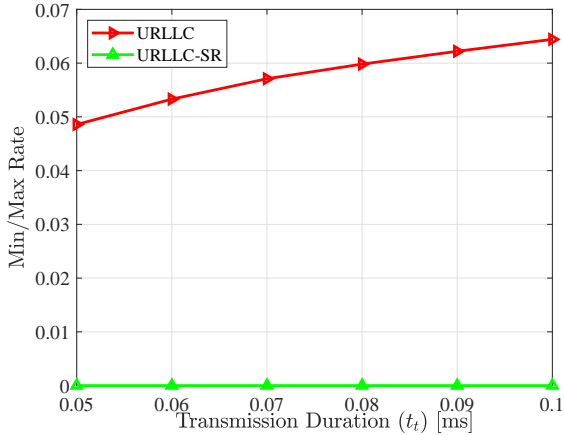}
	\caption{The ratio of the highest and lowest users rates versus $t_t$}
	\label{PGS-Figtime}
\end{figure}

\section{Conclusions} \label{sec4}
This paper has considered the joint design of transmit beamforming at the base station and RIS PREs
for RIS-aid multi-user URLLC. To guarantee the required quality-of-service in terms of downlink throughput in FBL regime while
maintaining computational tractability, we developed an algorithm, which invokes closed-form expressions at each iteration for generating a better point for the maximizing the geometric means of the users' rates (GM-rate). The algorithm has been supported  by  simulations.

\bibliographystyle{ieeetr}
\bibliography{TVT_final_21_07_22}

\end{document}